# Simulating Malicious Attacks on VANETs for Connected and Autonomous Vehicle Cybersecurity: A Machine Learning Dataset

Safras Iqbal, Peter Ball, Muhammad H Kamarudin, Andrew Bradley
*Oxford Brookes University*
*Oxford, United Kingdom*

## Abstract

Connected and Autonomous Vehicles (CAVs) rely on Vehicular Adhoc Networks with wireless communication between vehicles and roadside infrastructure to support safe operation. However, cybersecurity attacks pose a threat to VANETs and the safe operation of CAVs. This study proposes the use of simulation for modelling typical communication scenarios which may be subject to malicious attacks. The Eclipse MOSAIC simulation framework is used to model two typical road scenarios, including messaging between the vehicles and infrastructure - and both replay and bogus information cybersecurity attacks are introduced. The model demonstrates the impact of these attacks, and provides an open dataset to inform the development of machine learning algorithms to provide anomaly detection and mitigation solutions for enhancing secure communications and safe deployment of CAVs on the road.

**Keywords:** VANET, CAV, V2V, V2I, Cyber Attack, Replay Attack, Bogus information Attack, Intelligent Transport, Autonomous Vehicle

## 1. Introduction

The introduction of Connected and Autonomous Vehicles (CAVs) within the transportation industry promises improvements in road safety, and reductions in congestion and environmental pollution. Autonomous Vehicles (AVs) enable mobility benefits for elderly and disabled people who are unable to operate manually driven vehicles. However, to realise the full potential, CAVs will require wireless interaction with other vehicles and transport infrastructure to safely navigate various traffic conditions on public roads.

A Vehicular Adhoc NETwork (VANET) is a type of Mobile Adhoc Network for road environments that establish wireless communication between neighbouring vehicles and infrastructure within radio range. VANETs are realised using a combination of Roadside Units (RSU) and On-Board Units (OBU) [1]. This paper will focus on two modes of communication which are crucial to the safe operation of CAVs; Vehicle to Vehicle (V2V) - which enables vehicles to exchange driving information, and Vehicle to Infrastructure (V2I) - which defines how vehicles communicate with the infrastructure to exchange traffic-related attributes [2].

In V2I communication, Roadside Units (RSUs) are typically placed in radio range of each other so that communication can be established between them, expanding coverage of the VANET. RSUs can 'listen' to vehicular awareness data (e.g. speed, position, direction, acceleration) - transmitted in the form of safety messages, frequently broadcast between neighbouring vehicles. Updates about unexpected road events (i.e. breakdowns, accidents etc.) can be instantly communicated to authorities for immediate action e.g. to avoid significant road congestion. The RSU may also disseminate environmental awareness messages such as weather conditions, and speed restrictions.

However, despite the enormous potential of these connected vehicles, the nature of the automation and communication means that here are increased cybersecurity risks associated with the evolution of CAVs [3], [4], [5], [6], and thus it is vital to analyse the the risks within the vehicles, infrastructure, and communication links to identify the potential vulnerabilities and seek mitigation.

An emerging trend for analysing these threats is to model a road scenario and generate a dataset that contains vehicular awareness data and the V2V and V2I messaging [7], [8], [9]. These datasets can be used to develop Machine Learning (ML)-based solutions to mitigate against cyber security attacks. For example, [7] established a model and dataset for Distributed Denial of Service (DDoS). Using this dataset, mitigation techniques have been developed to counter the threat of DDoS attacks. Other work published a dataset for Sybil attack detection [8].

Many different types of potential cyber security attacks exist, including replay and bogus information attacks which present a potential threat to VANETs. In a replay attack, the attacker can listen to messages in the network and rebroadcast where required, e.g. impersonating an emergency vehicle with the aim of clearing the road ahead for personal benefit. Similarly, bogus information attacks can be imposed by vehicles that broadcast fabricated messages to the VANET and mislead other users [10]. These two types of attack have recently been identified as potential threats in a VANET [11]. However, to the best of our knowledge, there is no single unified approach to simulating these threats in a road environment and generating a dataset for mitigation techniques.

This paper describes a VANET simulation model that can run multiple realistic scenarios, emulating replay and bogus information attacks. The model is used to generate datasets that characterise the environment in which these attacks can take place, which can be used to allow researchers to train ML-based solutions to mitigate against these attacks.

This paper provides the following novel contributions:

- A methodology for modelling cyber-attacks on the VANET environment.
- A dataset that can be used by other researchers to train and evaluate new ML-based malicious threats detection and mitigation models for CAVs.

## 2. Related Work

*2.1 VANET security requirements*

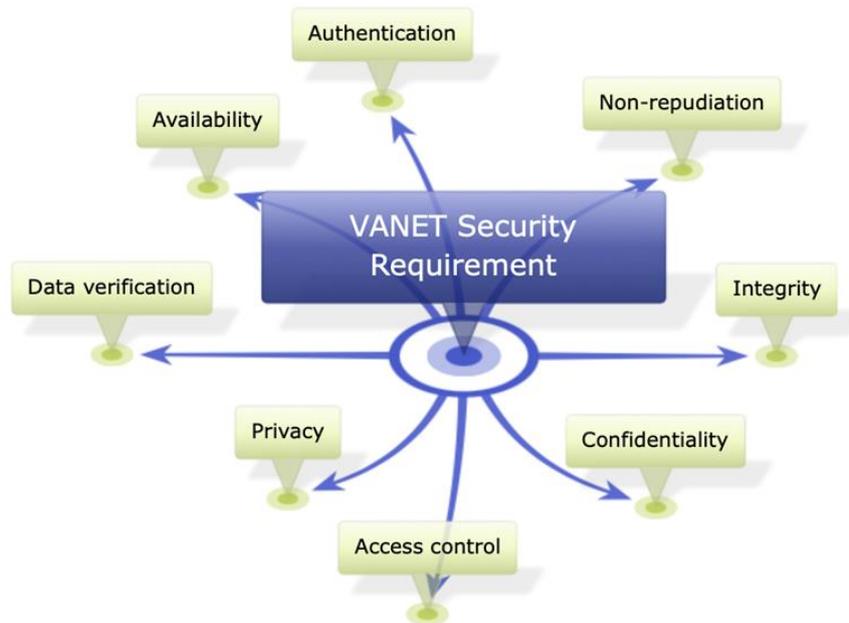

*Figure 1: VANET Security Requirements*

Security into any system usually leads to additional overhead. However, security is an essential element to keep VANET safe and trustable. Figure 1 highlights eight security features recognised as crucial aspects of VANET security [12], [13], [14].

VANET communication is established with secure authentication between nodes before exchanging any messages. However, in a replay attack, a legitimate user, authenticated through a digital signature, can turn into an attacker and perform a replay attack. Thus, the authentication requirement in Figure 1 is not satisfied by the cryptographic digital signature. These types of attackers are referred to as insider attackers.

Bogus information attacks pose a threat to the authentication and integrity of VANETs. This type of attack is possible when the attacker possesses a valid digital certificate for authentication and appears to be a legitimate user in the network. They can broadcast inaccurate and false messages to gain advantages. This attack can occur in many forms e.g. position falsification, speed falsification, and event falsification [11]. The attack could lead to re-routing and potentially a collision. As such, integrity and authentication need further attention, as per Figure 1.

*2.2 Vehicular communication technology*

VANET requires inter-vehicle communication; two main types of technologies are Dedicated Short Range Communication (DSRC) and 5G (C-V2X). This paper focuses upon DSRC technology, as it is the current technology in widespread deployment.

The Federal Communications Commission (FCC) allocated the 5.850 to 5.925 GHz band range with 75 MHz of the radio spectrum for DSRC to enhance road safety [15]. Wireless Access in Vehicular Environment (WAVE) is based on a communications stack with the DSRC IEEE 1609.x standards at the transport and network layers and the IEEE 802.11p standard at the MAC and physical layers.

DSRC uses Basic Safety Message (BSM), Probe Vehicle Data Message (PVD), Traveller Information Message (TIM), Map Data Message (MAP), Signal Phase and Timing Message (SPaT) - all defined by the SAE J2735 standard [15].

### 2.2.1 Vulnerabilities in vehicular communication

V2X messages are exchanged over the wireless medium and are vulnerable to cyber security threats. A malicious user could perform a jamming attack resulting in a denial of service (DoS). Furthermore, an eavesdropping attack could compromise confidentiality and lead to privacy concerns [16]. To mitigate against these vulnerabilities, both the IEEE and ETSI standardise and mandate the usage of a vehicular public key infrastructure (PKI). Therefore, vehicles in VANET use a digital signature to verify the sender's identity, and ensure the message is sent by a legitimate user. Although PKI can authenticate the sender, it does not guarantee the correctness of the message content [8]. As such, a vulnerability persists with insider malicious users who possess a valid digital certificate and are part of the VANET.

## 2.3 Attack simulations and data generation

Authors in [11] have modelled replay, message falsification, and a Sybil attack using the Veins simulation framework. Their work analysed the impact of these attacks in a VANET and proposed mitigation actions. They have proposed the use of encryption and authentication to mitigate against message falsification and replay attacks. However, it does not include a dataset. Other studies [17] suggest using certificateless aggregate signcryption scheme (CASS) solutions due to the overhead introduced by public key infrastructure (PKI)-based authentication. The CASS is designed to support anonymity and aggregation (two key requirements in safety warning systems).

Heijden et al. [9] have issued the publicly available Vehicular Reference Misbehaviour dataset (VeReMi); the purpose of this dataset is for the attack reproduction, comparison and analysis of ML-based VANET misbehaviour detection models. This dataset contains position falsification attacks with variable traffic density and attacker density of multiple scenario configurations. However, the evaluation workflow is non-interactive since attack detection is done post-simulation. Suppose input based on the detection algorithm for mitigation actions needs modification to the process. In [8], the VeReMi dataset has been extended by adding Sybil attack simulation data.

Gyawali et al. [16] developed a misbehaviour detection system (MDS) to detect anomalies in CAVs for two types of attacks. The first attack type is a false alert generation attack; a malicious vehicle can broadcast messages to the VANET, including emergency electronic brake lights, road hazard condition notifications, and collision warnings. The other type is a position falsification attack; the attacker may broadcast fake position data into the VANET. This may be used to initiate a Sybil attack in which multiple identities are created with fake locations. Based upon an ML model, the MDS system is trained using a dataset created from simulations of a false alert generation attack; for the position falsification attack, they use the VeReMi dataset [9]. However, the former dataset is not publicly available.

Determining threats in VANET and implementing countermeasures is crucial for the widespread deployment of VANET. In this paper we contribute simulations of replay attacks and bogus information attacks with realistic scenarios to demonstrate the attacks. The dataset will be helpful in training and validating machine learning based VANET adversary detection techniques.

## 3. Methodology

In this paper a method is presented to produce a dataset, and detect and mitigate replay attacks and bogus information attacks. A range of attributes is identified that can be used to detect the anomaly in a given scenario. For example, for replay attacks, message sequence numbers can be examined to check if the message generated at vehicle A (message ID XYZ) is the same as that received at vehicle B. The expected maximum time taken to deliver a message can also be taken into account. If this time is exceeded, this indicates the possibility of a replay attack. Likewise, for bogus information attack senders, a change in position is crucial for detecting a bogus information attack. For example, a fake accident report can be checked with other users in close proximity and if there are any inconsistencies an alert can be raised by the authority and RSUs so a plausibility check can be made with trusted RSUs and other VANET users in that area before taking any decision on the bogus information.

This study uses the Eclipse MOSAIC simulation framework to model the attacks. This framework is a collection of vehicular simulation tools as shown in Figure 2. OMNET++ is a core part of the MOSAIC that handles V2V and V2I communication. SUMO (Simulation of Urban Mobility) is the traffic simulation tool for 2D visualisation. The MOSAIC framework uses the SUMO core by connecting to the WebSocket visualiser tool to run on the web browser as shown in Figure 3. In the process of creating a scenario, we need to extract an area of a map. In our case, we select a region including roundabouts, a variable speed limit, and traffic lights. OpenStreetMap [18] was used to crop an area and export it as an "scenario.osm" file to extract the map.

The MOSAIC Scenario-Convertor tool is then used to create a database with all map-related tasks that MOSAIC will use. Then routes are added to the map database by providing start and end locations. This generates a folder structure where V2X applications can be loaded into the scenario, add vehicles and RSUs.

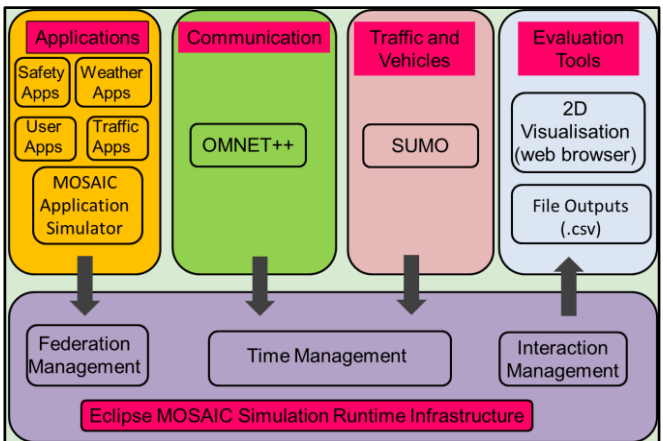

Figure 2: Eclipse MOSAIC simulator architecture

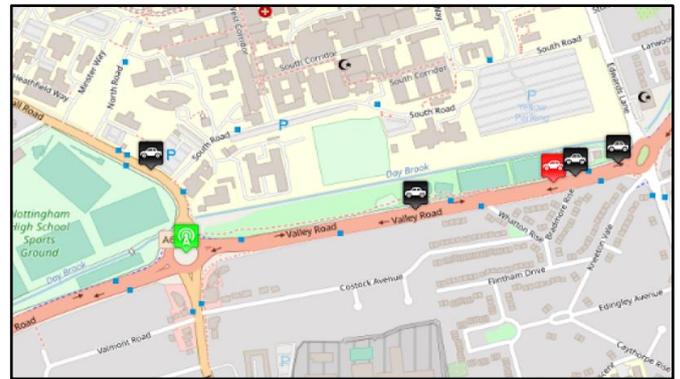

Figure 3: MOSAIC runtime visualiser

### 3.1 *V2X Application Development*

V2X applications for each attack scenario have been written in java, and a compiled jar file is attached to the scenario to establish V2X communication. Corporate Awareness Messages (CAM) are used to perform V2V and V2I communication in the attack scenarios.

## 3.2 Process of Dataset Generation

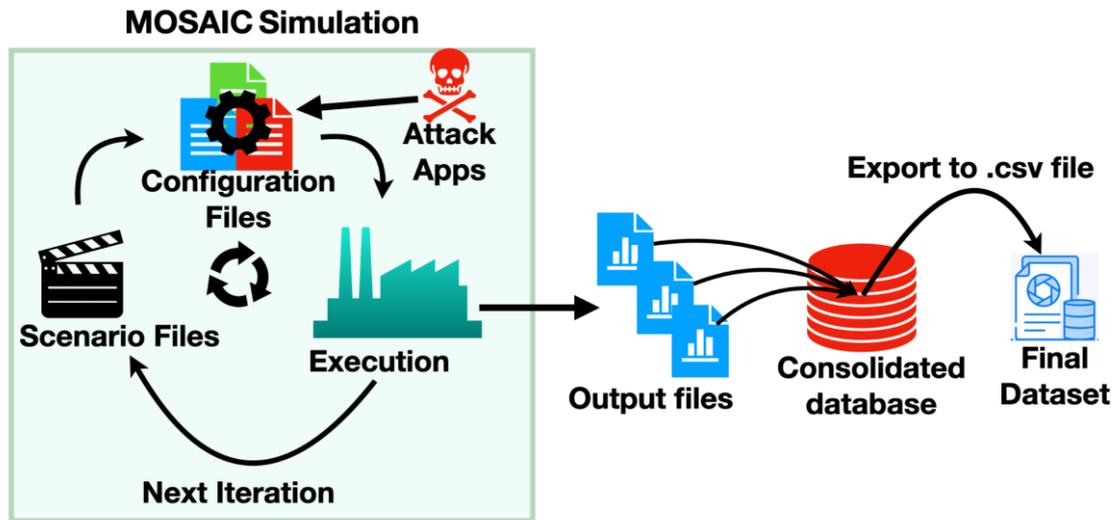

Figure 4: Dataset generation framework architecture

Figure 4 depicts the dataset generation process. The first step is to create a scenario (as explained above). Then attach the application based on the attack type and set parameters in the mapping configuration file. The next step is to execute each scenario for the number of iterations required for each setting. Every iteration generates a subset of output files in CSV format. Subsequently, the output files are consolidated into a database for export to the final dataset. The same process generates three different datasets, one without attack, the second set with a replay attack and a third with a bogus information attack.

## 3.3 Dataset samples

| Message Type | Time | Name | Speed | Heading | Latitude | Longitude | Altitude | DistanceDriven | LongitudinalAcceleration | Slope | Stopped | RouteId | Connection.Id | LaneIndex | BlinkerRight | BlinkerLeft | BrakeLight |
|---|---|---|---|---|---|---|---|---|---|---|---|---|---|---|---|---|---|
| VEHICLE_UPDATES | 4000000000 | veh_0 | 1.848357785 | 153.8736876 | 52.99163946 | -1.169330223 | 0 | 1.848357785 | 0.8483577853 | 0 | FALSE | 1 | 30806029_34133( | 0 | TRUE | FALSE | FALSE |
| VEHICLE_UPDATES | 5000000000 | veh_0 | 2.830734692 | 154.0742482 | 52.9916163 | -1.169312763 | 0 | 4.679092478 | 0.9823769071 | 0 | FALSE | 1 | 30806029_34133( | 0 | TRUE | FALSE | FALSE |
| VEHICLE_UPDATES | 6000000000 | veh_0 | 3.438154875 | 154.0742482 | 52.99158817 | -1.169291556 | 0 | 8.117247353 | 0.6074201827 | 0 | FALSE | 1 | 30806029_34133( | 0 | TRUE | FALSE | FALSE |
| VEHICLE_UPDATES | 7000000000 | veh_0 | 4.130656012 | 154.0742482 | 52.99155438 | -1.169266078 | 0 | 12.24790336 | 0.6925011366 | 0 | FALSE | 1 | 30806029_34133( | 0 | TRUE | FALSE | FALSE |

Figure 5: sample vehicle attributes

Figure 5 shows a sample structure of the message data, which periodically records the status of every vehicle in the simulation. The details of each field are given in Table I below.

| Vehicle update fields ||
|---|---|
| **Attribute Name** | **Description** |
| **Time** | Simulation time in nanoseconds (ns) |
| **Name** | Vehicle Id where the message received from |
| **Speed** | Speed of the vehicle in metre per seconds (m/s) |
| **Heading** | Vehicle headings direction in degrees clockwise |
| **Latitude** | Decimal degree format of latitude |
| **Longitude** | Decimal degree format of longitude |
| **Altitude** | Altitude in metres from sea level |
| **DistanceDriven** | Distance driven during the simulation (m) |
| **LongitudinalAcceleration** | Acceleration of the vehicle in m/s2 |
| **Slope** | Slope in degrees |
| **Stopped** | Boolean value for stopped True/False |

| RouteId | Number of ruote in the scenario in integer |
|---|---|
| Connection ID | Unique road ID constructed with 2 edge ids |
| LaneIndex | Lane used, 0 most left to 1,2 .. |
| BlinkerRight | Boolean value for signal blinker True/False |
| BlinkerLeft | Boolean value for signal blinker True/False |
| BrakeLight | Boolean value for BrakeLight True/False |

*Table 1: Vehicle Update field dictionary*

Figure 6 shows a sample structure of the V2X message data, which records every time a vehicle or RSU disseminates a V2X message to the VANET. The details of each field are discussed in Table II below.

| Message Type | Time | Type | MessageId | SourceName | Latitude | Longitude | Altitude | Destination.Type | IPv4Address | AdhocChannelId |
|---|---|---|---|---|---|---|---|---|---|---|
| V2X_MESSAGE_TRAN | 1676632782 | Cam | 0 | rsu_0 | 52.987078 | -1.15962 | 0 | AD_HOC_TOPOCAST | /255.255.255.255 | CCH |
| V2X_MESSAGE_TRAN | 4028794307 | Cam | 3 | veh_0 | 52.99163946 | -1.169330223 | 0 | AD_HOC_TOPOCAST | /255.255.255.255 | CCH |
| V2X_MESSAGE_TRAN | 4676632782 | Cam | 4 | rsu_0 | 52.987078 | -1.15962 | 0 | AD_HOC_TOPOCAST | /255.255.255.255 | CCH |
| V2X_MESSAGE_TRAN | 5028794307 | Cam | 5 | veh_0 | 52.9916163 | -1.169312763 | 0 | AD_HOC_TOPOCAST | /255.255.255.255 | CCH |

*Figure 6: sample V2X message transmission*

| V2X message transmission fields ||
|---|---|
| Attribute Name | Description |
| Time | Simulation time in nanoseconds (ns) |
| Type | Message type transmitted E.g CAM/DENM |
| MessageId | Message sequence number |
| SourceName | Vehicle ID who transmitted the message |
| Latitude | Decimal degree format of latitude |
| Longitude | Decimal degree format of longitude |
| Altitude | Altitude in metres from sea level |
| Destination.Type | Method of radio usage |
| IPv4Address | IP address of the node |
| AdhocChannelId | Message transmitted channel e.g CCH, SCH |

*Table 2: V2X message transmission field dictionary*

| Message Type | Time | Type | MessageID | ReceiverName | Receive Signal Strength |
|---|---|---|---|---|---|
| V2X_MESSAGE_RECEPTION | 6030422107 | Cam | 7 | veh_1 | 0 |
| V2X_MESSAGE_RECEPTION | 7029913786 | Cam | 9 | veh_1 | 0 |
| V2X_MESSAGE_RECEPTION | 8014783295 | Cam | 11 | veh_0 | 0 |
| V2X_MESSAGE_RECEPTION | 8030098398 | Cam | 12 | veh_1 | 0 |

*Figure 7: sample V2X message received*

Figure 7 shows a sample structure of the V2X message data, which records every time a vehicle or RSU receives a V2X message from the VANET. The details of each field are discussed in Table III below.

| V2X message transmission fields ||
| Attribute Name | Description |
| --- | --- |
| Time | Simulation time in nanoseconds (ns) |
| Type | Message type received e.g CAM/DENM |
| MessageID | Message sequence number |
| ReceiverName | Vehicle ID who received the message |
| ReceiveSignalStrength | Signal Strength when receiving the message |

*Table 3: V2X message reception field dictionary*

### 3.4 Attack Simulation

#### 3.4.1 Replay attack

A replay attack is generally a pre-recorded set of legitimate messages transmitted through a V2X communication. For example, in an emergency braking scenario, a set of beacons are disseminated to neighbouring vehicles as part of safety messages. Therefore, other vehicles can warn the driver or apply the braking automatically upon receiving this message to prevent an accident as part of an advanced driving assistance application. However, an attacker could use these pre-recorded messages and can re-transmit them into, for example, a harmonised platoon system or a high speed and dense traffic environment to cause a collision in the moving traffic. A replay attack is also referred to as a playback attack [19].

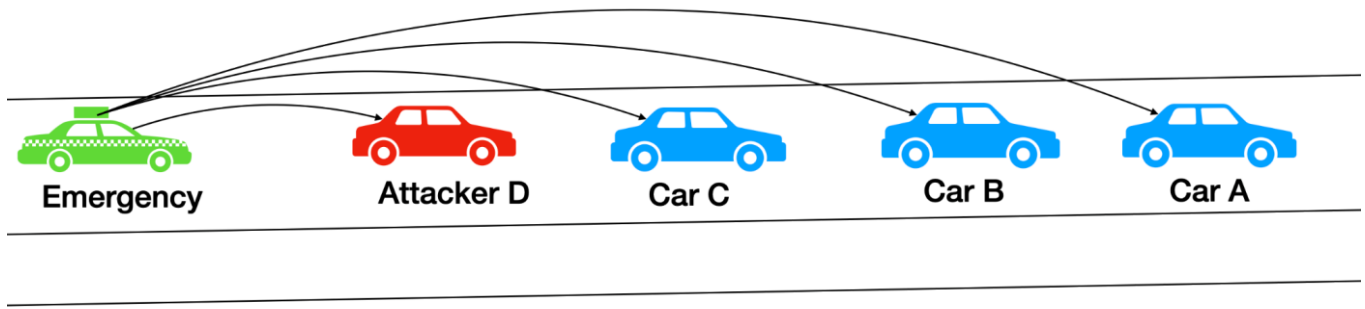

*Figure 8: Replay attack setup*

*Scenario 1: Emergency Vehicle Replay Attack*

An attack scenario simulated to demonstrate the replay attack on CAVs. The method assumes that when an emergency vehicle approaches the four vehicles should stop and resume when the emergency vehicle passes by as in Figure 8. As such, preceding vehicles must stop when emergency vehicles reach 100 metres in proximity. This represents a typical real world scenario. However, if we assume that vehicle number 4 is an attacker and records the messages sent by an emergency vehicle and then retransmits them to the VANET. Thus, the attacker can gain a personal advantage on the road to get past other cars.

#### 3.4.2 Bogus Information attack

A bogus information attack refers to the transmission of false information/events to the VANET as shown in Figure 9. The attacker's objective is to gain personal advantage, for example by diverting traffic to alternative routes by sending false traffic information or road accidents information [20].

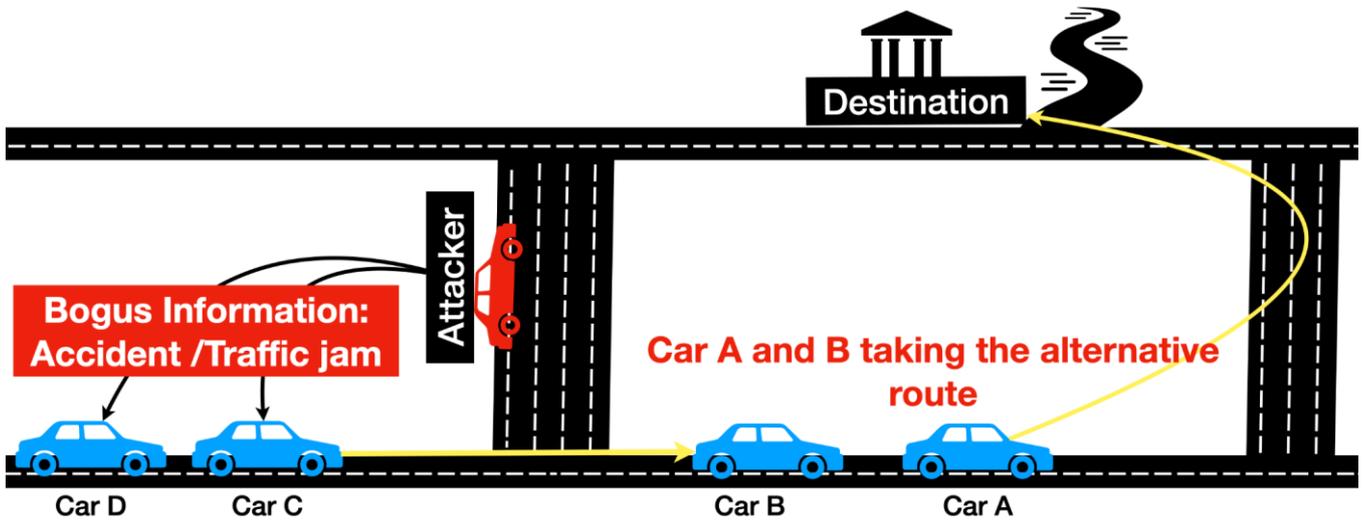

*Figure 9: Bogus Information attack setup*

*Scenario 2: Bogus Information Attack with false announcement*

In this scenario, a number of vehicles travel from point A to point B, and alternative routes are available based on the map used in this scenario. The first vehicle is an attacker who attempts to broadcast a bogus message into the VANET to indicate that there is an accident along the current route. This directs the other vehicles onto other routes and allows the attacker to travel along a clear route.

## 4. Results and Discussion

4.1 *Results*

4.1.1   Scenario 1: Replay attack

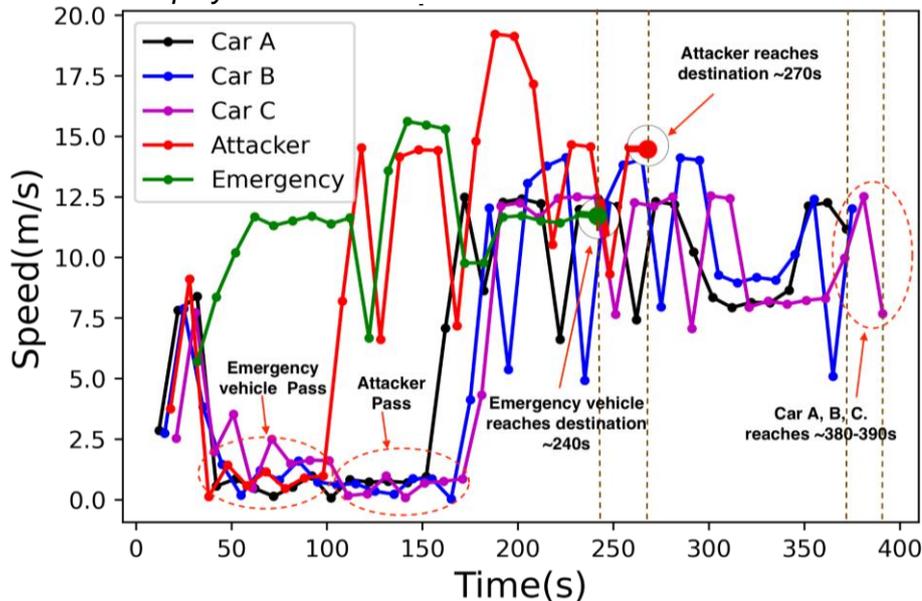

*Figure 10: Replay attack speed chart. Note that vehicles A, B and C stop to allow the emergency vehicle to pass, while the 'attacker' vehicle (D) uses a replay attack to pass effortlessly through traffic by impersonating the emergency vehicle.*

Figure 10 shows the effect of a replay attack that impersonates an emergency vehicle. The last vehicle to enter the scenario is the emergency vehicle, reaching the destination first at around 240s. The second last vehicle is the attacker labelled as the attacker (vehicle D from Figure 8) records the message from the approaching emergency vehicle and replays the V2x messages to stop the cars ahead. Vehicles A, B, C are the victims' vehicles that reach the destination with a delay caused by a malicious user.

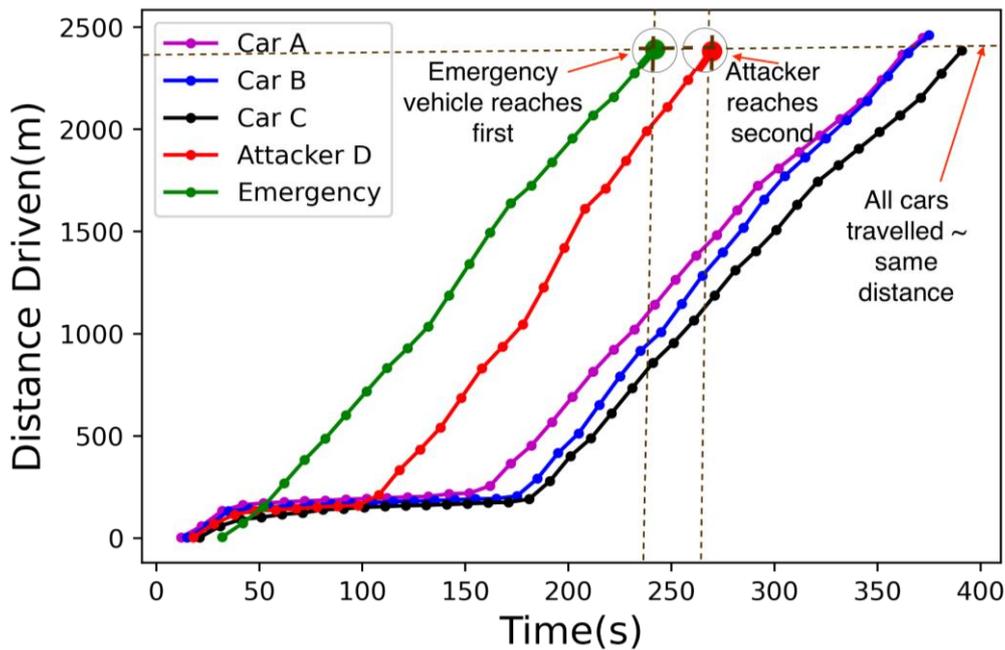
Figure 11: Replay attack distance driven chart

Emergency vehicles reach the destination first and are followed by the attacker. Figure 11 shows the distance driven from start to end. All the vehicles travelled the same distance.

### 4.1.2 Scenario 2: Bogus Information attack

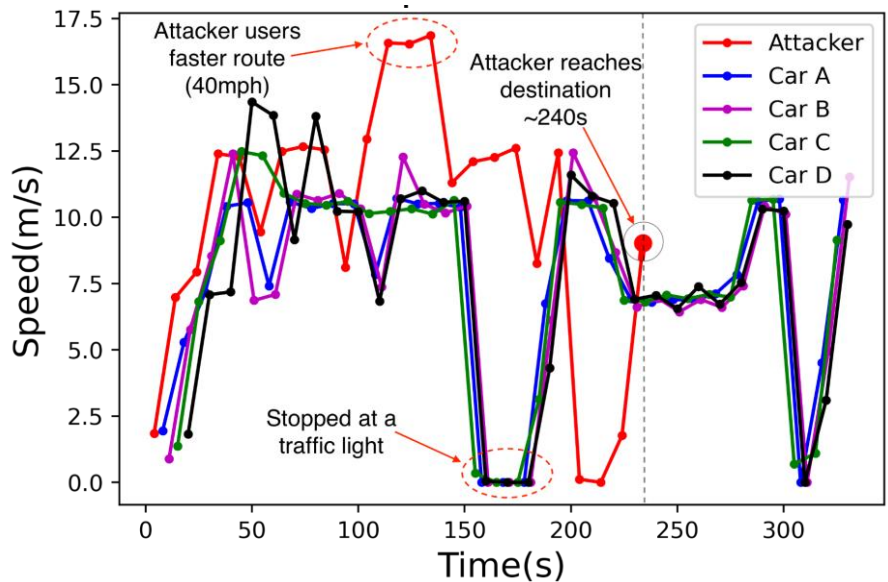
Figure 12: Bogus Information attack speed chart

As shown in Figure 12, speed data is examined from the bogus information attack data. As scenario two explained in the previous section, the leading vehicle is the attacker who disseminates bogus information to the network. The following four vehicles received the message via decentralised environment notification message (DENM) to indicate their route ahead is interrupted with an accident. Therefore, the V2X application suggests an alternative route to reach the destination. As per the speed data, the attacker reached the destination around 240s, well ahead of the other four cars, and other vehicles reached around 340s as per the simulation. This is due to others taking the alternative route.

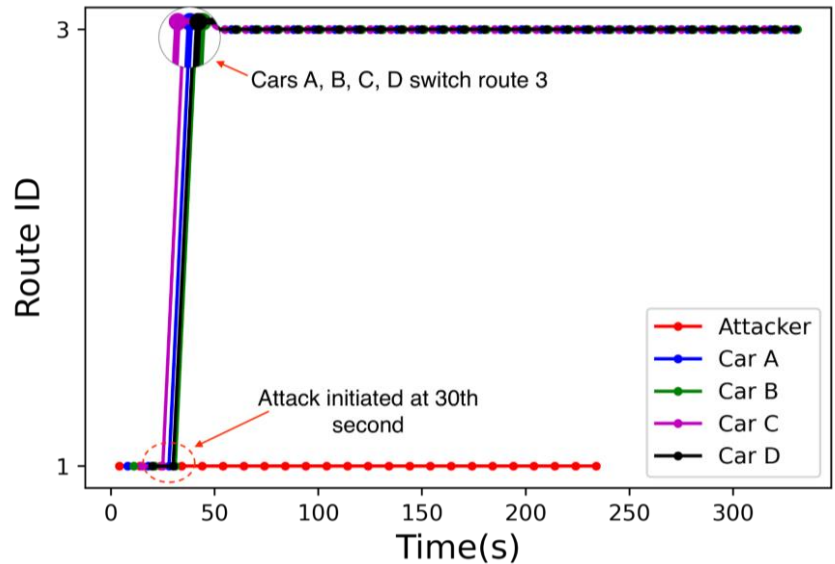

*Figure 13: Bogus Information attack route chart*

Figure 13 indicates that all five vehicles were in route 1 and the attack took place at 30s, then the other vehicles were rerouted via route 3 till the destination. Also this indicates the attacker reached the destination ahead of others.

The link to the dataset is given below.[1]

### 4.2 *Discussion*

The information in the dataset presented in this paper provides information that can be used for identifying two types of attack: replay attacks and bogus information attacks. An extensive range of data parameters are included such as speed, GPS position, heading, acceleration, distance driven, timestamp etc. Time sequence data and message sequence numbers can be used to investigate replay attacks and inconsistencies in location information from different users in close proximity, and may be used as the basis for identifying replay and bogus information attacks. Further work will focus upon the use of the data from these scenarios for developing algorithms to prevent such attacks from compromising the VANET.

## 5. Conclusion

This paper presents a novel methodology for using simulation to model cybersecurity attacks in Vehicular Ad-hoc Networks. The MOSAIC simulation framework was used to simulate traffic flow, vehicle-to-vehicle and vehicle-to-infrastructure communications, with a custom module developed to emulate malicious attacks (including replay and bogus information attacks) - with the resulting effects on vehicle behaviour demonstrated and detection methods outlined.

An open dataset[1] for developing and training machine learning-based mitigations against cybersecurity attacks has been generated using the model, enabling other researchers to develop threat detection and mitigation techniques to enhance the cybersecurity of connected and autonomous vehicles, and thus accelerate safe deployment on the world's roads.

---

[1] https://drive.google.com/drive/folders/1dDrt1K90B4zn1zi6WT3ZsZ4GstPwMLWg

**References**

[1] Offor, P., 2012. Vehicle ad hoc network (vanet): Safety benefits and security challenges. Available at SSRN 2206077.
[2] Sheikh, M. S., Liang, J. and Wang, W. (2019) 'A survey of security services, attacks, and applications for vehicular ad hoc networks (vanets)', Sensors, 19(16), pp. 3589.
[3] Gupta, R., Tanwar, S., Kumar, N. and Tyagi, S., 2020. Blockchain-based security attack resilience schemes for autonomous vehicles in industry 4.0: A systematic review. Computers & Electrical Engineering, 86, p.106717.
[4] El-Rewini, Z., Sadatsharan, K., Selvaraj, D.F., Plathottam, S.J. and Ranganathan, P., 2020. Cybersecurity challenges in vehicular communications. Vehicular Communications, 23, p.100214.
[5] Khan, S.K., Shiwakoti, N., Stasinopoulos, P. and Chen, Y., 2020. Cyber-attacks in the next-generation cars, mitigation techniques, anticipated readiness and future directions. Accident Analysis & Prevention, 148, p.105837.
[6] Kim, K., Kim, J.S., Jeong, S., Park, J.H. and Kim, H.K., 2021. Cybersecurity for autonomous vehicles: Review of attacks and defence. Computers & Security, p.102150.
[7] Kadam, M.N. and Sekhar, K.R., 2021. Machine Learning Approach of Hybrid KSVN Algorithm to Detect DDoS Attack in VANET. Machine Learning, 12(7).
[8] Kamel, J., Wolf, M., van der Hei, R.W., Kaiser, A., Urien, P. and Kargl, F., 2020, June. Veremi extension: A dataset for comparable evaluation of misbehavior detection in vanets. In ICC 2020-2020 IEEE International Conference on Communications (ICC) (pp. 1-6). IEEE.
[9] Heijden, R.W., Lukaseder, T. and Kargl, F., 2018, August. Veremi: A dataset for comparable evaluation of misbehavior detection in vanets. In International Conference on Security and Privacy in Communication Systems (pp. 318-337). Springer, Cham.
[10] Ghaleb, F.A., Zainal, A., Maroof, M.A., Rassam, M.A. and Saeed, F., 2019. Detecting bogus information attack in vehicular ad hoc network: a context-aware approach. Procedia Computer Science, 163, pp.180-189.
[11] Lastinec, J. and Keszeli, M., 2019, June. Analysis of realistic attack scenarios in vehicle ad-hoc networks. In 2019 7th International Symposium on Digital Forensics and Security (ISDFS) (pp. 1-6). IEEE.
[12] Afzal, Z. and Kumar, M., 2020. Security of vehicular Ad-hoc networks (VANET): A survey. In Journal of Physics: Conference Series (Vol. 1427, No. 1, p. 012015). IOP Publishing.
[13] Engoulou, R.G., Bellaïche, M., Pierre, S. and Quintero, A., 2014. VANET security surveys. Computer Communications, 44, pp.1-13.
[14] Engoulou, R.G., Bellaïche, M., Pierre, S. and Quintero, A., 2014. VANET security surveys. Computer Communications, 44, pp.1-13.
[15] J. B. Kenney, "Dedicated Short-Range Communications (DSRC) Standards in the United States," in Proceedings of the IEEE, vol. 99, no. 7, pp. 1162-1182, July 2011, doi: 10.1109/JPROC.2011.2132790.
[16] Gyawali, S. and Qian, Y., 2019, May. Misbehavior detection using machine learning in vehicular communication networks. In ICC 2019-2019 IEEE International Conference on Communications (ICC) (pp. 1-6). IEEE.
[17] Yang, Y., Zhang, L., Zhao, Y., Choo, K.K.R. and Zhang, Y., 2022. Privacy-Preserving Aggregation-Authentication Scheme for Safety Warning System in Fog-Cloud Based VANET. IEEE Transactions on Information Forensics and Security.
[18] Haklay, M. and Weber, P., 2008. Openstreetmap: User-generated street maps. IEEE Pervasive computing, 7(4), pp.12-18.
[19] M. A. Al-shareeda, M. Anbar, I. H. Hasbullah, S. Manickam, N. Abdullah and M. M. Hamdi, "Review of Prevention schemes for Replay Attack in Vehicular Ad hoc Networks (VANETs)," 2020 IEEE 3rd International Conference on Information Communication and Signal Processing (ICICSP), 2020, pp. 394-398, doi: 10.1109/ICICSP50920.2020.9232047.
[20] Stępień, K. and Poniszewska-Marańda, A., 2021. Security Measures with Enhanced Behavior Processing and Footprint Algorithm against Sybil and Bogus Attacks in Vehicular Ad Hoc Network. Sensors, 21(10), p.3538.